# Effects of substituting Se with Te in the FeSe compound: structural, magnetization and Mössbauer studies.

R.W. Gómez[1], V. Marquina[1], J.L. Pérez-Mazariego[1], R. Escamilla[2], R. Escudero[2], M. Quintana[1], J.J. Hernández-Gómez[1], R. Ridaura[1] and M.L. Marquina[1].

[1]Facultad de Ciencias, Universidad Nacional Autónoma de México. Circuito Exterior, CU.
México D.F., 04510
México
[2]Instituto de Investigaciones en Materiales, Universidad Nacional Autónoma de México, CU.
Apartado Postal 70-360, México D.F., 04510
México

**Abstract**

Polycrystalline samples of $FeSe_{1-x}Te_x$ (x = 0.00, 0.25, 0.50, 0.75 and 1.00) were synthesized by solid-state reaction to study the effects of substituting Se with Te in the system. The magnetization properties of the resulting compounds were investigated and the crystallographic structures of the samples analyzed through X-ray diffraction. Mössbauer spectroscopy was used to determine the ionic state of the Fe ions and the hyperfine fields. The magnetic susceptibility curves of the samples with x = 0.25, 0.50 and 0.75 show superconducting behavior. The lattice parameters and the cell volume increase monotonically with increasing Te concentration and the Mössbauer spectra reveal the absence of internal magnetic hyperfine fields.

Keywords: Iron based superconductors, Mössbauer effect, X-ray diffraction, magnetization.

PACS: 74.25.Fy, 7425.Ha 74.62.Bf, 74.70.Dd.

## 1. Introduction

The discovery of the La[O$_{1-x}$F$_x$]FeAs superconductors [1], prompted vigorous research focusing on this system and the related, and simpler, binary compound FeSe [2-7 and references therein]. A lot less attention, however, has been paid to the Se-substituted ternary compound Fe(Se$_{1-x}$Te$_x$) [8-13].

In order to explain superconductivity in these systems, several proposals have been made: a) Neutron-scattering results, performed in LaFeFAsO, reveal the existence of a spin-density wave antiferromagnetic order [14] indicating that magnetic fluctuations could play a role in the pairing mechanism between charge carriers in the F-doped compound. b) The enhancement of the critical superconducting temperature (T$_c$) with applied pressure, suggests that the structural characteristics of LaO$_{1-x}$F$_x$FeAs [15] and of FeSe [16,17] may modify the concentration of charge carriers. c) The structural phase transition observed by McQueen et al [18] at 90 K in the Fe$_{1.01}$Se superconductor, and by Margadonna et al at 70 K [19], could be related to the formation of a charge density wave to explain the observed characteristics.

The substitution of Se by Te introduces a chemical pressure into the structure of the Fe(Se$_{1-x}$Te$_x$) system that may have similar effects to a mechanical pressure. To study this and other possible structural changes, X-ray diffraction studies were made, including Rietveld analysis of the X-ray patterns. Also, the magnetic properties (superconductivity) of the system were investigated by zero field cooling (ZFC) and field cooling (FC) magnetic susceptibility measurements. In addition, the presence of Fe ions allows the use of Mössbauer spectroscopy to provide information of the electric and magnetic environment around the Fe ions, and also about the possible structural changes in the Fe($Se_{1-x}Te_x$) system.

## 2. Experimental procedures

Polycrystalline samples of Fe($Se_{1-x}Te_x$), with x = 0.0, 0.25, 0.50, 0.75 and 1.0, were synthesized from Fe (Baker 99.99%), Se (Alfa Aesar 99.5%) and Te (Alfa Aesar 99.99%) mixed in stoichiometric amounts, sealed in an evacuated quartz tube and heated at 700 °C for seven days, after which they were cooled to room temperature. The resulting samples were ground to obtain fine powders to perform the measurements mentioned above.

Phase identification of the powder samples was performed with a Siemens D5000 X-ray diffractometer using Cu-$K_\alpha$ radiation and a Ni filter. Intensities were measured at room temperature in 0.02° steps, in the 6° - 130° 2θ range. The crystallographic phases were identified by comparison with the X-ray patterns of the JCPDS database. In order to perform the analysis, the spectra were refined using the Rietica Rietveld program, v 1.71 with multi-phase capability [20].

Zero field cooling (ZFC) and field cooling (FC) magnetic susceptibility versus temperature (χ-T) measurements were taken with a Quantum Design superconducting quantum interference device (SQUID) based magnetometer, in the temperature range of 2 K to 30 K on 120 mg encapsulated samples.

Thin absorbers were made with powder of the samples to record room temperature Mössbauer spectra in transmission geometry with a constant acceleration spectrometer, using a $^{57}$Co Mössbauer source in rhodium matrix. All the spectra were fitted using the Recoil 1.05 [21] program and the isomer shifts are quoted with respect to iron.

## 3. Results and discussion

*3.1 XRD analysis and crystal structure*

The powder X-ray diffraction patterns obtained for Fe(Se$_{1-x}$Te$_x$) with x = 0.00, 0.25, 0.50, 0.75 and 1.00 are shown in Figure 1. The spectrum of the sample without Te (x = 0.00) corresponds to the FeSe tetragonal (ICDD nº 85-0735) PbO-type common structure, although minute traces of hexagonal-NiAs-type FeSe Achavalite (ICDD n° 65-91284) are observed. A splitting in the diffraction peaks for the x = 0.25 composition is observed, indicating the presence of two structural phases. The fact that in this case the sample presents superconductivity around 12 K (see Figure 4) implies that one of the phases must have the FeSe$_{1-x}$Te$_x$ composition, but with x ≠ 0.25. Actually, from the result of the Rietveld analysis (Table I), only 62.8 % of the ternary compound is formed, giving then a composition FeSe$_{0.843}$Te$_{0.157}$, together with the FeTe phase. This has also been observed by other groups [9, 11], and can be a consequence of the larger radius of the Te ions, which inhibits inter-atomic diffusion in the FeSe lattice. The x = 0.50 and 0.75 samples present only small amounts (~5%) of the hexagonal NiAs-type FeSe and of the FeTe$_2$ (ICDD n° 89-2090) phases, respectively. Lastly, in the sample with x = 1.00 shows mainly the FeTe tetragonal PbO-type (ICDD nº 89-4077) and minute traces of FeTe$_2$. The X-ray diffraction patterns were Rietveld-fitted using the space group P4/*nmm* (nº 227) considering the different Te compositions. The fitted pattern for the x = 0.50 composition is shown in Figure 2. The Rietveld refinement results are summarized in the Table 1, including a list of selected bond lengths and bond angles.

    Figure 3 shows the evolution of the *a*-axis and *c*-axis lattice parameters and unit cell volume for the Fe(Se$_{1-x}$Te$_x$) system as a function of the Te content (x). Using the calculated Te concentration mentioned above, the crystal lattice parameters and the

volume cell increase almost linearly as the Te concentration is increased. It is important to note that, due to the larger ionic radius of Te, as compared to that of Se, the Fe-Fe and Fe-(Se,Te) bond-lengths also increase. The evolution of $T_c$ with the Te concentration goes through a maximum at x = 0.5, revealing that the relation between superconductivity in this system and its structure requires more thought.

3.2 Magnetic susceptibility

The ZFC and FC magnetic susceptibility curves of the samples with negative magnetization (x = 0.25, 0.50 and 0.75), in the 2 K to 30 K range, are shown in Figure 4. We do not show the curves for FeSe and FeTe, given their lack of superconductivity. The sample with x = 0.5 exhibited the highest critical temperature, $T_c$, of about 13 K. Figure 5 presents another susceptibility curve for this sample in the 2 K to 300 K interval. The abnormal behavior seen in the 13 K to about 150 K interval of the ZFC curve is probably associated with a tetragonal to triclinic structural transformation [10] or to a tetragonal to orthorhombic transformation [18]. Also, the positive values of the suceptibility could be due to traces of iron oxide in the sample. Other researchers have also observed anomalies in the susceptibility curve of the similar system β-FeSe around 100 K [7, 9].

The Meissner and shielding characteristics of our measurements allow determining the amount of the superconducting proportion of the samples. The x = 0.5 sample presents complete Meissner behavior of about 85 %.

3.3 Mössbauer spectroscopy

Room temperature Mössbauer spectra of all samples were recorded and are shown in Figure 6 and their Mössbauer parameters (quadrupole splitting ΔQ and isomer shift IS) and relative populations are shown in Table 2, together with the corresponding relative populations calculated from the Rietveld refinement of the X-ray diffraction patterns; the table also shows the results of a quadrupole splitting calculation as described in the next paragraph. The spectra of the x = 0.00, 0.50, 0.75 and 1.00 samples were fitted with two quadrupole doublets. The one with x = 0.25 requires three quadrupole doublets, revealing the presence of more than two phases. In Table 2, the tetragonal phases are labeled as A

(FeSe), A' (FeSe$_{1-x}$Te$_x$) and A" (FeTe), the hexagonal FeSe phase as B and the FeTe$_2$ phase as C.

A simple point charge calculation for the electric field gradients and their corresponding quadrupole splittings was performed as a guide to assign the observed doublets to the different ionic configurations around the Fe ions. The stoichiometry of the samples, the observed Mössbauer parameters and the absence of hyperfine magnetic sextets in the spectra suggests that the ionic state of iron must be Fe$^{2+}$ in a low spin configuration (S = 0), so the calculations were made with this assumption. The calculations were carried out to the second nearest neighbors in virtue of the small difference of bond lengths between the Fe-Se or Fe-Te ions and the Fe-Fe ions (see Table 1). First, the electric field gradient components in the principal axes system were calculated with

$$V_{ij} = \sum q_k \left(3 x_{i,k} x_{j,k} - r_k^2 \delta_{i,j}\right) r_k^{-5},$$

where $q_k$ are the ionic charges around the Fe ion (the origin), $x_{ik}$ their coordinates and $r_k$ their distances to the Fe ion. Secondly, the magnitudes of the quadrupole splittings are calculated as:

$$\Delta Q = \frac{1}{2} qQV_{zz} \left(1 + \frac{\eta^3}{3}\right)^2 (1 - \gamma_\infty)$$

where Q is the quadrupole moment of the nucleus, $\eta = (V_{xx} - V_{yy})/V_{zz}$ is the asymmetry parameter and $\gamma_\infty$ is the Sternheimer factor. The recent value of Q = 0.16 b, reported by Martínez-Pinedo et al [22] was used. The results of the calculations are listed in Table 2, with no consideration of the Sternheimer factor, in which the following structures are included: tetragonal β-FeSe, hexagonal α-FeSe, tetragonal FeSe$_{1-x}$Te$_x$, FeTe and FeTe$_2$.

Comparing the results of these calculations with the experimental values of the quadrupole splittings, it is clear that the only configuration that can be associated with the smallest quadrupole splitting (ΔQ ~ 0.1 mm/s) is the hexagonal FeSe phase, that the one with the highest one (ΔQ ~ 0.9 mm/s) corresponds to the FeTe$_2$ phase and that the one with the intermediate value (ΔQ ~ 0.3 mm/s) is the one to be associated with the FeSe$_{1-x}$Te$_x$ phase.

Conclusions

The studies performed on the FeSe$_{1-x}$Te$_x$ system (X-ray diffraction, magnetic susceptibility and Mössbauer spectroscopy) bring about consistent and complementary results: a) No signs of magnetism are observed in the susceptibility curves nor in the Mössbauer spectra. b) The results of the compositions of the samples from the Rietveld X-ray diffraction refinement are consistent with those inferred from Mössbauer spectroscopy. c) The samples with the nominal composition x = 0.25, 0.50 and 0.75 are superconducting; however, all our attempts to synthesize the sample with x = 0.25 produced two main phases (FeSe$_{0.843}$Te$_{0.157}$ and FeTe) and one minor hexagonal FeSe phase, which explains the presence of three quadrupole doublets in its Mössbauer spectrum. The above composition is inferred by both X-ray diffraction studies and Mössbauer spectroscopy. d) The magnetic susceptibility curve taken in the 2 K to 300 K range shows an anomaly around 100 K which is probably associated with a structural change similar to that observed in the FeSe compound. e) The lattice parameters and volume cell increase almost linear as Te concentration is increased.

Acknowledgments

We would like to acknowledge Victor Hugo Ortiz for sample preparation. This work was partially supported by DGAPA-UNAM, projects IN-110808, IN-111408 and IN-227208.

Table Captions.

Table 1 Structural parameters obtained from the Rietveld refinement analysis of the Fe(Se$_{1-x}$Te$_x$) at 295 K. Space group: *P 4/n m m* (# 129). Atomic positions: Fe: 2a (0, 0, 0); Se: 2c (0, ½, 0.26); % of impurity in the phase; B (in A$^2$) and N are the isotropic thermal and occupancy parameters, respectively.

Table 2. Observed Mössbauer parameters: Isomer shift (IS) respect to iron, quadrupole splitting-($\Delta$Q), site population percent from Mössbauer (Mossb %) and X-ray (XRD %) results and calculated quadrupole splitting ($\Delta$Q$_{cal}$).

Figure captions.

Figure 1. X-ray diffractograms of the five samples. The observed impurities are marked as * : hexagonal FeSe, + : FeTe and º : $FeTe_2$.

Figure 2. Rietveld refinement for the x = 0.50 sample. Experimental spectrum (dots), calculated pattern(continuous line), their difference (middle line) and the calculated peak positions (bottom).

Figure 3. Variation of the lattice parameters and cell-volume with the Te content (x).

Figure 4. ZFC and FC magnetic susceptibility curves of the superconducting samples.

Figure 5. ZFC magnetic susceptibility of the x = 0.5 sample in the 2 K to 300 K interval.

Figure 6. Mössbauer spectra of the $FeSe_{1-x}Te_x$ system, with x = 0.0 (FeSe), 0.25, 0.50, 0.75 and 1.0 (FeTe).

# Tables

Table 1

| x= | 0.0 | 0.25 | 0.50 | 0.75 | 1.0 |
|---|---|---|---|---|---|
| a(Å) | 3.7752(1) | 3.7872(2) | 3.7913(2) | 3.8129(2) | 3.8266(1) |
| c(Å) | 5.5268(2) | 5.6492(3) | 5.9784(3) | 6.1500(3) | 6.2935(1) |
| V(Å$^3$) | 78.77(2) | 81.03(2) | 85.93(3) | 89.41(3) | 92.15(1) |
| %FeSe$_{1-x}$Te$_x$ | 82.8(2) | 62.8(2) | 91.6(3) | 93.10(3) | 94.9(3) |
| %FeTe | - | 35.4(2) | - | - | - |
| %FeSeHex | 18.2(1) | 1.8(1) | 8.4(9) | - | - |
| %FeTe$_2$ | - | - | - | 6.9(1) | 5.1(2) |
| Bond length (Å) | | | | | |
| Fe - Fe: 4 | 2.670(3) | 2.678(2) | 2.681(3) | 2.696(3) | 2.706(2) |
| (Se/Te) - Fe: 4 | 2.373(2) | 2.397(3) | 2.451(2) | 2.488(3) | 2.559(2) |
| Bond angle (º) | | | | | |
| (Se/Te)-Fe-(Se/Te) | 105.4(3) | 112.1(2) | 113.7(3) | 114.4(3) | 115.0(2) |
| Fe-(Se/Te)-Fe | 68.5(2) | 67.9(3) | 66.3(3) | 65.6(3) | 65.0(2) |
| Fe — B (Å$^2$) | 0.30(4) | 0.37(3) | 0.42(3) | 0.49(3) | 0.49(2) |
| Fe — N | 0.98(2) | 0.97(2) | 0.98(2) | 0.96(2) | 0.96(1) |
| (Se/Te) — B (Å$^2$) | 0.71(2) | 0.67(1) | 0.82(3) | 0.49(2) | 0.49(4) |
| (Se/Te) — N | 1.01(2) | 1.03(2) | 1.01(2) | 1.03(1) | 1.03(1) |
| $R_{wp}$ (%) | 22.1 | 18.1 | 18.2 | 17.2 | 10.1 |
| $R_p$ (%) | 17.4 | 14.2 | 14.3 | 13.2 | 7.5 |
| $R_{exp}$ (%) | 17.5 | 16.6 | 16.4 | 16.1 | 6.3 |
| $\chi^2$ | 1.2 | 1.1 | 1.1 | 1.1 | 1.6 |

Table 2

| x | Doublet | IS | ΔQ | Mössb. % | XRD % | Structure | ΔQ$_{cal}$ |
|---|---|---|---|---|---|---|---|
| **0.00** | A | 0.45(17) | 0.29(18) | 79.3(46) | 82.8 | FeSe (tetragonal) | 0.1152 |
|  | B | 0.13(33) | 0.09() | 20.7(65) | 18.2 | FeSe (hexagonal) | 0.026 |
| **0.25** | A' | 0.44(18) | 0.29 | 63.9(12) | 62.8 | FeSe$_{0.843}$Te$_{0.157}$ | 0.1174 |
|  | A" | 0.47(16) | 0.36(11) | 33.3(86) | 35.4 | FeTe | 0.1288 |
|  | B | -0.28() | 0.1() | 2.8(16) | 1.8 | FeSe (hexagonal) | 0.0260 |
| **0.50** | A' | 0.45(50) | 0.29 | 97.5(21) | 91.6 | FeSe$_{0.50}$Te$_{0.50}$ | 0.1262 |
|  | B | 0.19(74) | 0.1() | 2.5(11) | 8.4 | FeSe (hexagonal) | 0.0260 |
| **0.75** | A' | 0.45(13) | 0.29(19) | 94.5(37) | 96.4 | FeSe$_{0.25}$Te$_{0.75}$ | 0.1274 |
|  | C | 0.18(61) | 0.84(12) | 5.5(32) | 3.6 | FeTe$_2$ | 0.2128 |
| **1.00** | A" | 0.46(11) | 0.32(17) | 93.6(69) | 94.9 | FeTe | 0.1288 |
|  | C | 0.29(34) | 0.99(73) | 6.4(64) | 5.1 | FeTe$_2$ | 0.2128 |

Figure 1

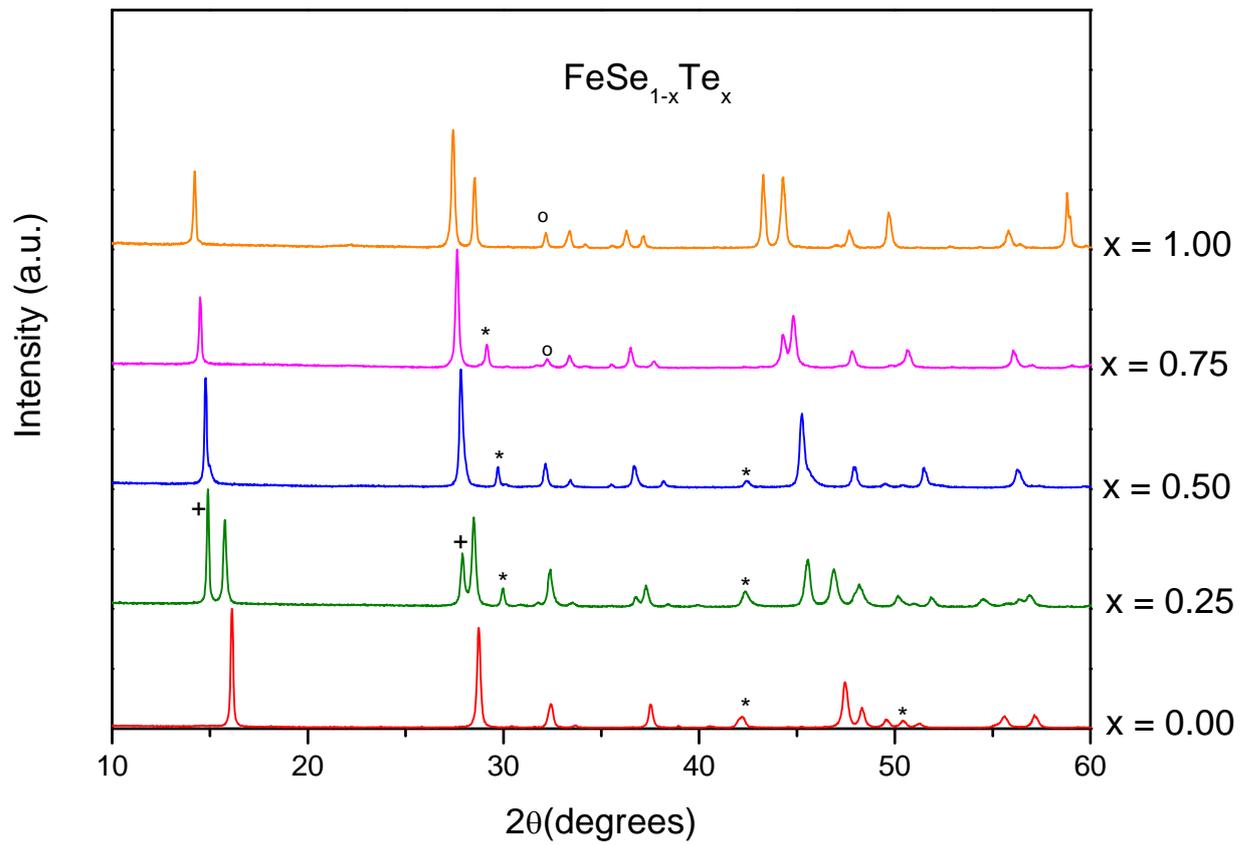

Figure 2

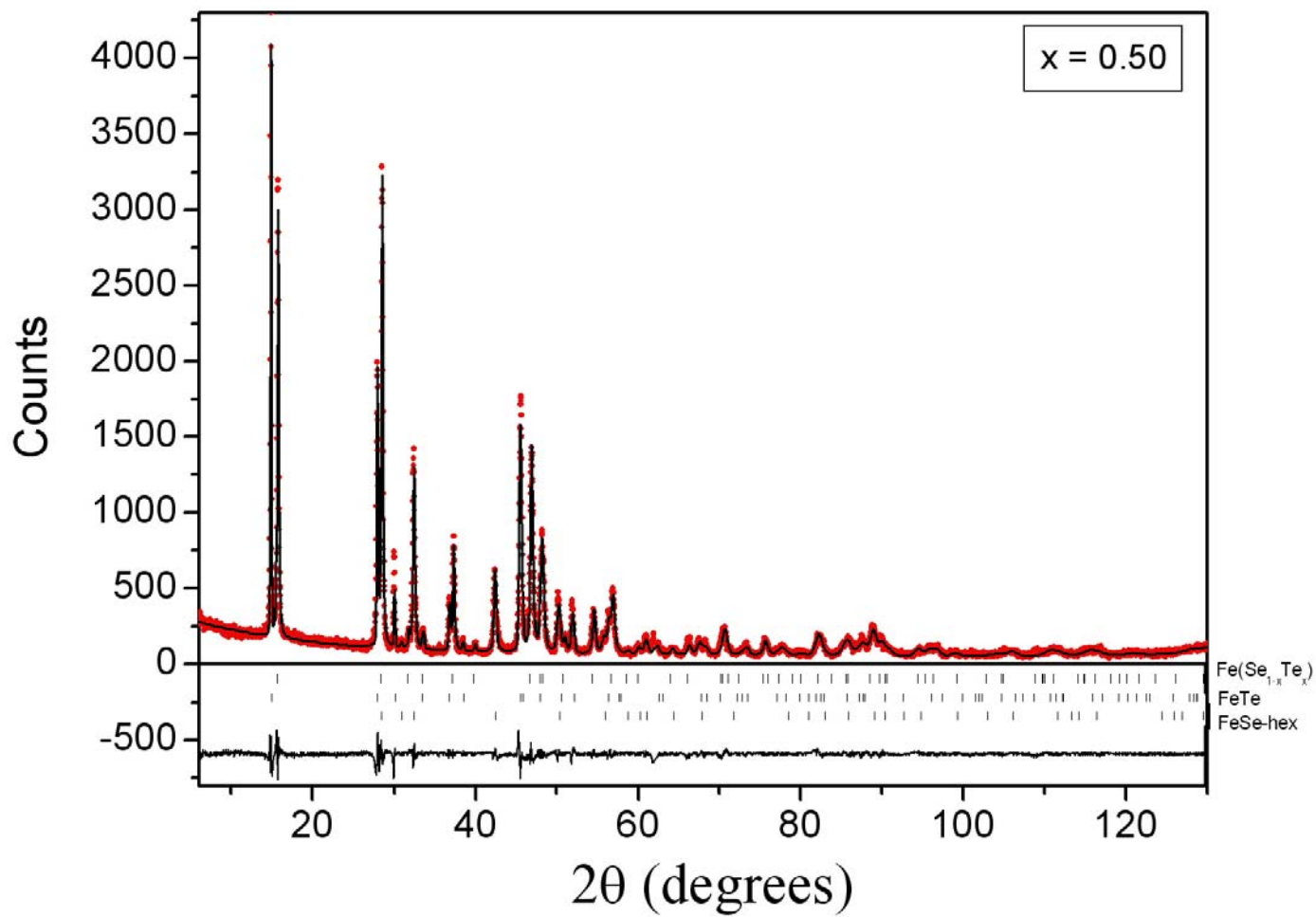

Figure 3

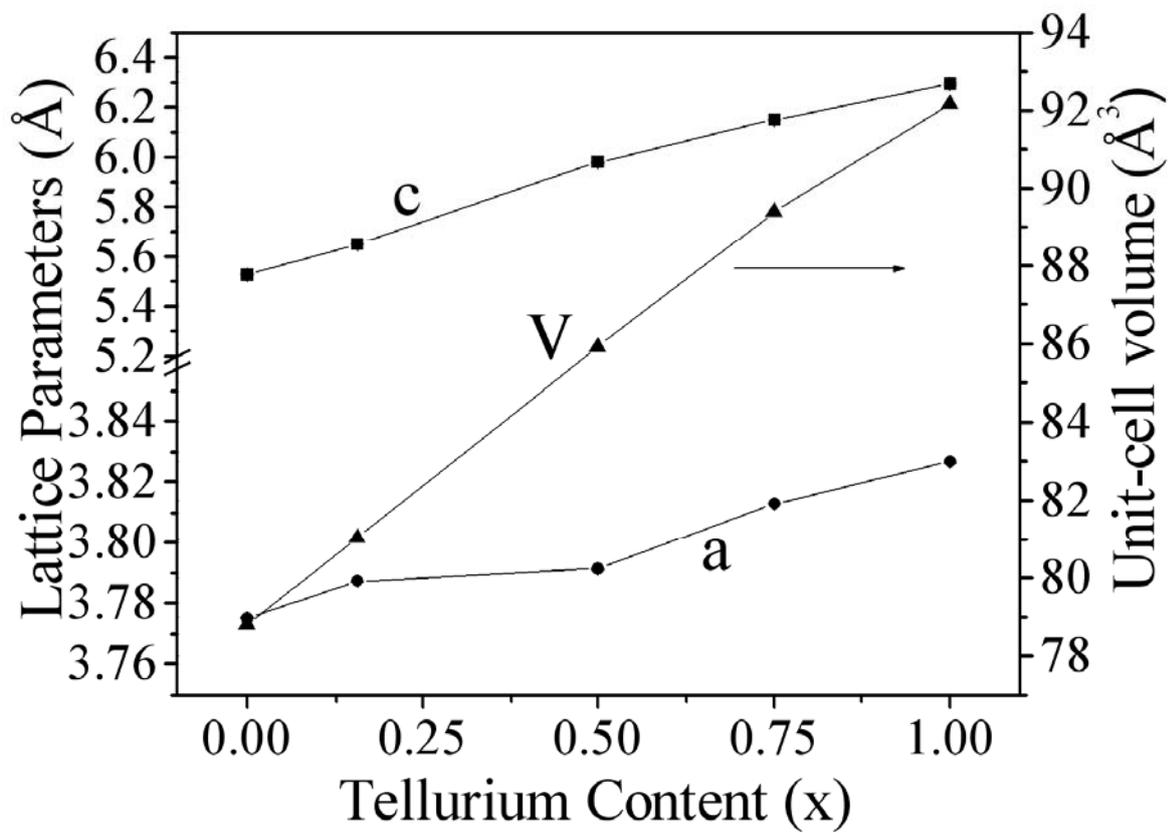

Figure 4

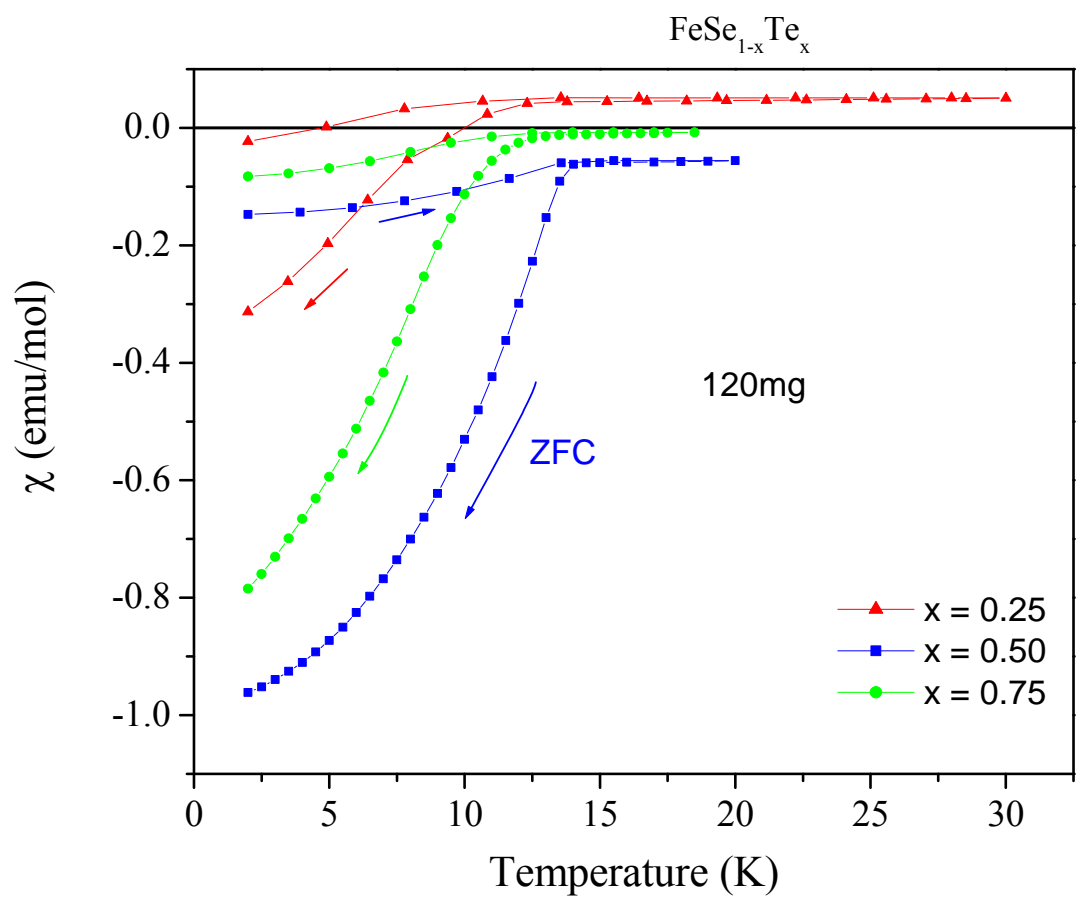

Figure 5

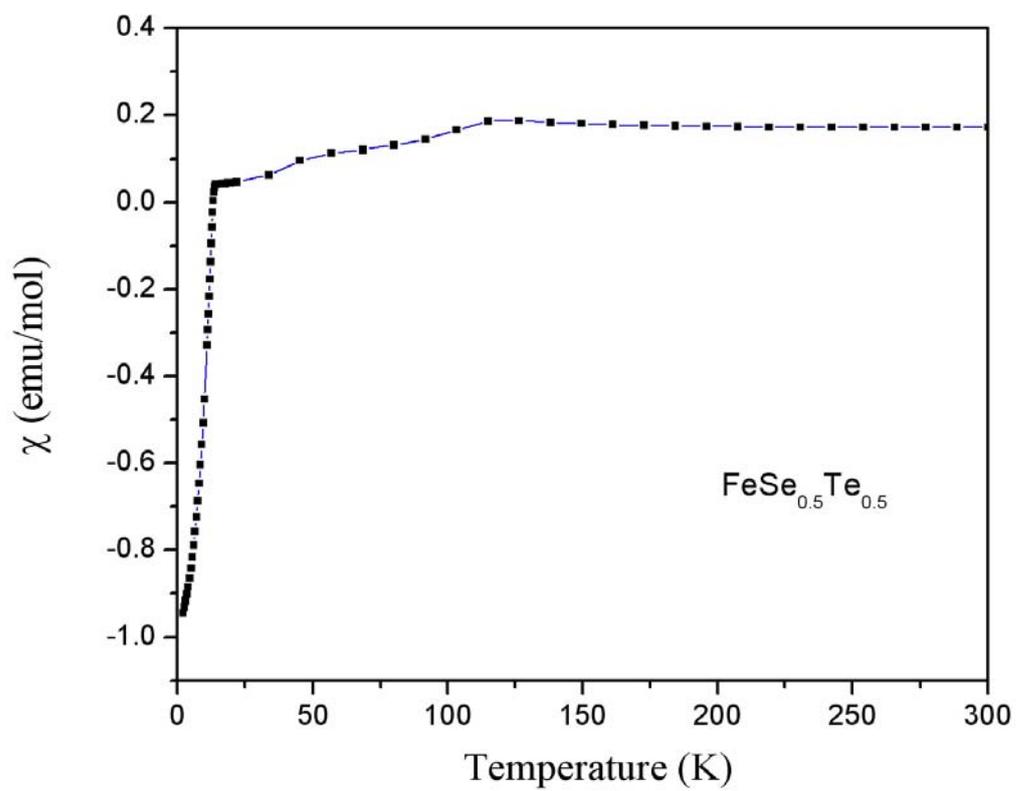

Figure 6

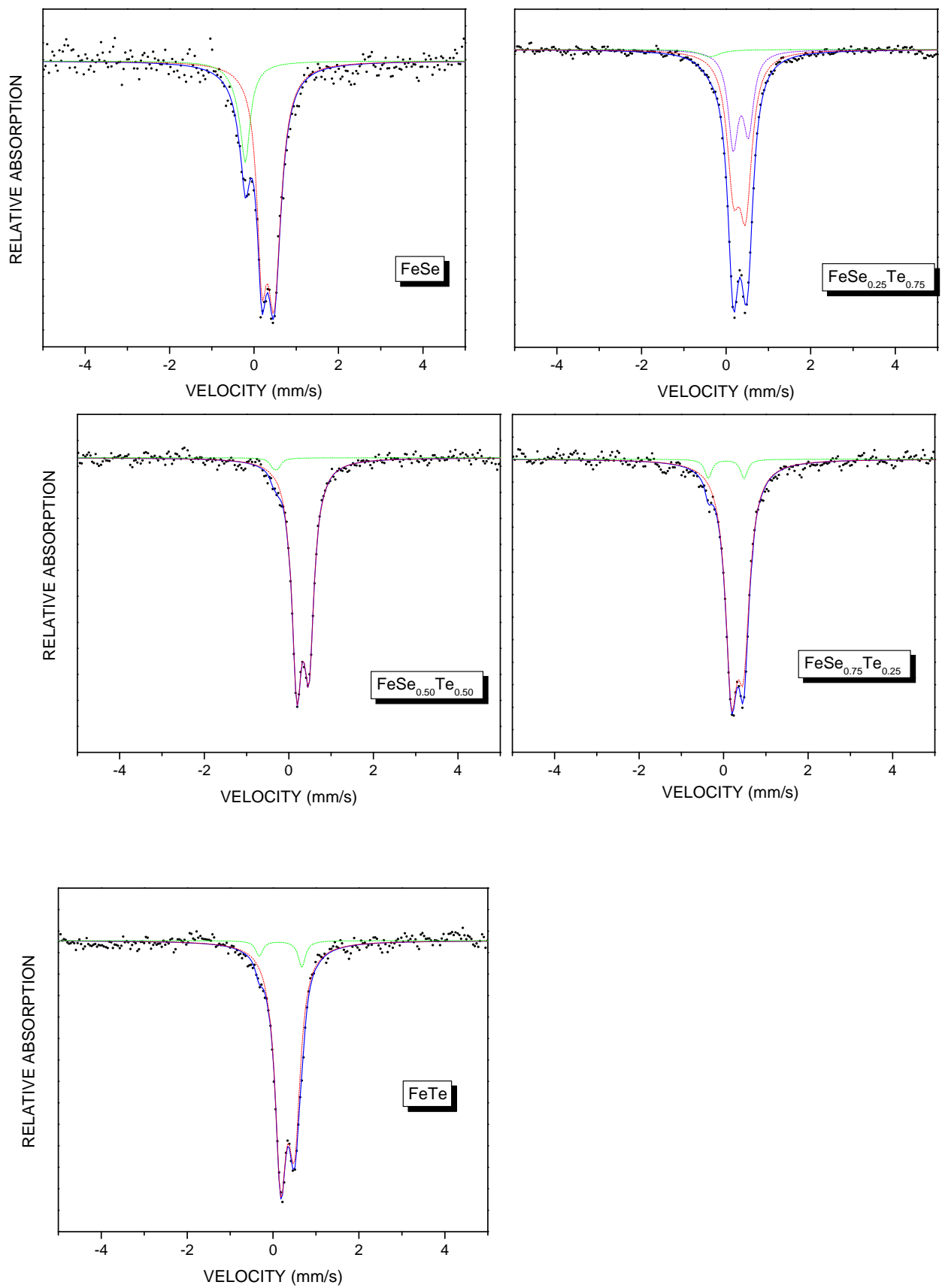